\documentstyle[prl,aps,multicol,epsf]{revtex}
\tighten
\begin{document}
\draft
\title{
Adiabatic Charge Pumping in Almost Open Dots
}
\author{I.L. Aleiner$^{1}$ and A.V. Andreev$^{2}$}
\address{$^{1}$Department of Physics and Astronomy, 
SUNY at Stony Brook, Stony Brook, NY 11794\\
$^{2}$
Institute for Theoretical Physics,
UCSB, Santa Barbara, CA 9106-4030
}
\maketitle

\begin{abstract}
We consider adiabatic charge transport
through an almost open quantum dot.
We show that the charge transmitted in one cycle is quantized in the
limit of vanishing temperature and one-electron mean level spacing in
the dot. The explicit analytic expression for the pumped charge
at finite temperature is obtained for spinless electrons.
The pumped charge is produced
by both non-dissipative and dissipative currents.
The latter are responsible for the corrections to charge quantization
which are expressed through the conductance of the system.
\end{abstract}

\begin{multicols}{2}

Adiabatic charge pumping occurs in a  system subjected
to a slow periodic perturbation. Upon the completion of
the cycle, the Hamiltonian of the system returns to its
initial form, however, a finite
charge may be transmitted  through some cross-section of the system.
The natural question  is what is the value of this charge transmitted
through the system during one cycle, $Q$. This question does not
have a universal answer. Thouless\cite{Thouless83} showed
that for certain one dimensional systems with a gap in the
excitation spectrum in the
thermodynamic limit the charge $Q$ is quantized.
Such quantized charge pumping could be of practical importance
as a standard of electric current\cite{Niu}. The accuracy of
charge quantization depends on the degree of adiabaticity of the
process.

The practical attempts at creating a  quantized electron pump
use a different approach based on the phenomenon of
Coulomb blockade\cite{Likharev,Devoret}.
In this kind of devices, one uses several single
electron transistors (SET) connected in series to increase
the accuracy of charge quantization. At least two
SET's are necessary to obtain a non-zero charge
transfer during one cycle.

Recently, another family of the Coulomb blockade devices was
demonstrated\cite{review} -- semiconductor based quantum dots. The
advantage of these devices is the possibility of changing 
not only the gate
voltage and thus the average electron number in the dot but also the
conductance of the quantum point contacts (QPC's) separating the
dot from the leads. By doing so, one can traverse from almost
classical Coulomb blockade to the completely
open dot where the effects of the charge 
quantization are diminished.

Having those semiconductor structures in mind, we theoretically study
 in this Letter the following setup for the adiabatic quantum pump.
The device is depicted in Fig.~1 and consists
of a quantum dot connected to two reservoirs labeled by $\alpha=\pm 1$
by one channel QPC's characterized by reflection amplitudes $r_\alpha(t)$
and capacitively coupled to the metallic gate G. The gate
potential is characterized by the number of electrons $N$ which
minimizes the electrostatic energy of the dot.
Experiments with a similar setup were reported in
Ref.~\cite{Kouwenhoven}.
\narrowtext{
\begin{figure}
\begin{center}
\epsfxsize=7cm
\epsfbox{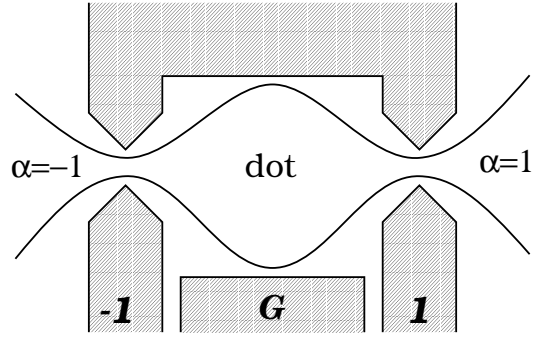}
\end{center}
\caption{ Schematic drawing of a quantum dot electrostatically
defined on a surface of a two dimensional electron gas.
The dot is connected to two leads by single channel
QPC's labeled by  $\alpha$.
The voltages on the gates ``$G$'' and ``$\pm 1$'' 
determine respectively
the average electron number in the dot, $N(t)$, and the
reflection amplitudes, $r_{\pm 1}(t)$, in the QPC's.}
\label{fig:1}
\end{figure}
}
We show below that the charge adiabatically transfered in one cycle is
quantized in the limit of zero temperature and zero mean level
spacing $\Delta$. For spinless electrons it is given by
\begin{equation}
\frac{Q}{e} =\frac{1}{2\pi i} \oint \frac{dz}{z}, \quad
 \label{charge}
z(t)=\sum_\alpha r_{\alpha}(t)e^{i \alpha\pi N(t)}.
\label{z}
\end{equation}
We assume that during the cycle the system does not go through the
degeneracy point $z=0$. The degree of adiabaticity of the
process depends on the proximity to this degeneracy point.

To illustrate this quantization qualitatively,
let us first consider  the following
trivial limit of the pumping cycle: i) the right contact ($\alpha=+1$)
is completely pinched off, ii) the gate voltage is changed from $0$ to
$N_0$,
iii) the left contact ($\alpha=-1$) is adiabatically closed.
In this process the average charge on the dot changes from
$N_0$ to the nearest integer $n_0$ (If $N_0$ is not a
degeneracy point, which is assumed, $n_0$ is unique),
iv) the right contact is opened, and the gate voltage is changed from
$N_0$ back to $0$, v) the left contact is opened. As a result of this
cycle the total charge transfered from left to right is an integer
$n_0$.
The non-trivial statement is that even for contacts which stay {\em
weakly }
reflecting at all times during the cycle the total transfered charge is
still quantized and is given by Eq.~(\ref{charge}).

We assume that the electrons in the leads are non-interacting,
and the interaction between the electrons on the dot is
described by the standard model of Coulomb blockade for strong
tunneling\cite{Matveev95,Furusaki95}, see Sec. 4.3 of
Ref.~\cite{Aleiner97} for the discussion of the applicability of this
model. If the dot is connected to the leads of by single channel
QPC's the system can be treated within the
one-dimensional effective Hamiltonian\cite{Matveev95,Furusaki95}
(we put $\hbar=k_B=1$ everywhere)
\end{multicols}
\widetext
 \begin{eqnarray}
 \hat{H}(t)&=&iv_F \sum_{\alpha }
\int_{-\infty}^{+\infty}dx\left( \psi^{ \dagger}_{L,\alpha}(x)
\partial_x \psi_{L,\alpha}(x) -\psi^{ \dagger}_{R,\alpha}(x)
\partial_x \psi_{R,\alpha}(x)\right)
+ v_F \sum_\alpha \left( r_\alpha(t)  \psi^{ \dagger}_{R,\alpha}(0)
\psi_{L,\alpha}(0)+h.c.\right) \nonumber \\
& +& E_c \left( \sum_{\alpha} \int^{+\infty}_{0}dx
:\psi^{ \dagger}_{L,\alpha}
(\alpha x) \psi_{L,\alpha}(\alpha x)
+ \psi^{ \dagger}_{R,\alpha}(\alpha x)\psi_{R,\alpha}(\alpha x):
+N(t)\right)^2.
 \label{Heff}
 \end{eqnarray}
\begin{multicols}{2}

The first two terms in Eq.~(\ref{Heff}) describe the
non-interacting electrons
in the left ($\alpha=-1$) and the right ($\alpha=+1$) QPC
respectively: the first term is the
linearized version of the kinetic energy ($v_F$ is the Fermi velocity)
and the second term corresponds to backscattering in the QPC's. We
restrict
ourselves to the case of small reflection amplitudes, $r_{\alpha}(t)\ll
1$.
Finally, the last term describes the effect of the Coulomb
blockade in the dot, and $E_C$ is its charging energy. For
simplicity, we explicitly consider the case of spinless electrons,
the results for the electrons with spins will
be presented at the end of the paper. The only difference of our
Hamiltonian from those considered
before\cite{Matveev95,Furusaki95,Aleiner97}
resides in its time dependence.
Since upon the completion of the cycle the total charge of the dot
returns
to its original value the integrated current can be calculated
through any cross-section of the system. For convenience we
take half the sum of the currents flowing through the left
and right point contacts:
\begin{equation}
\hat{I}=\frac{ev_F}{2} \sum_\alpha \left( :
\psi^\dagger_{L,\alpha}(0)\psi_{L,\alpha}(0)-
\psi^\dagger_{R,\alpha}(0)\psi_{R,\alpha}(0):\right),
\label{current}
\end{equation}
where $e$ is the electron charge.

Similarly to Ref.~\cite{Furusaki95}, the Hamiltonian (\ref{Heff})
can be bosonized according the rules
\begin{equation}
\psi_{L/R, \alpha}(x)=\frac{\hat{\eta}_{\alpha}}
{\sqrt{2\pi\lambda}}
\exp\left(\pm i \frac{\hat{\phi}^I_{L/R,\alpha}(x)+ \alpha
\hat{\phi}^C_{L/R, \alpha}(x) }{\sqrt{2}}\right)
\label{rules}
\end{equation}
where $\hat{\eta}_{\alpha}$ are Majorana fermions
($\hat{\eta}_{\alpha}= \hat{\eta}_{\alpha}^\dagger$,
$\{\hat{\eta}_{\alpha},\hat{\eta}_{\alpha'}\}=
2\delta_{\alpha, \alpha'}$), whereas the scale
$\lambda$ characterizes the large momentum cut-off
and is of the order of the Fermi wavelength.

Instead of the left and the right modes in Eq.~(\ref{rules}) it
is convenient to introduce the even and odd modes
$\hat{\phi}^{I,C}_{\pm}$ as
\begin{equation}
\hat{\phi}^{i}_{L/R}(x)= \frac{\hat{\phi}^{i}_{+}(\pm
x)\pm\hat{\phi}^{i}_{-}(\pm x)}
{\sqrt{2}}; \quad i=I, C.
\label{evenodd}
\end{equation}
The bosonic operators $\hat{\phi}^{I,C}_{\pm}(x)$
satisfy the following commutation relations
\begin{eqnarray}
&&\left[ \hat{\phi}^{i}_{+}(x), \hat{\phi}^{j}_{+}(y)\right]=
\left[ \hat{\phi}^{i}_{-}(x), \hat{\phi}^{j}_{-}(y)\right]=-i\pi
{\rm sgn} (x-y) \delta_{ij}  \nonumber\\
&&\left[ \hat{\phi}^{i}_{-}(x), \hat{\phi}^{j}_{+}(y)\right]=
i\pi\delta_{ij}
; \quad i,j =I, C.
\label{relations}
\end{eqnarray}
The last of Eqs.~(\ref{relations}) ensures the correct anticommutation
relation between left
and right moving fermions, however, this subtlety will not be important
for the problem in hand.

The odd modes $\hat{\phi}^{i}_-$ are decoupled from the rest of
the Hamiltonian and do not contribute to the current
(\ref{current}), hence we can omit them.
The relevant part of the Hamiltonian (\ref{Heff})
acquires the following form
\begin{eqnarray}
&&\hat{H}(t) =\frac{v_F}{4\pi}\!\sum_{i=I,C}\!
\int_{-\infty}^{+\infty}\!\!\!\! dx
\left( \frac{\partial \hat{\phi}^{i}_{+}(x)}{\partial x} \right)^2
+E_c\left[\frac{\hat{\phi}^{C}_{+}(0)}{\pi}- N\right]^2 \nonumber \\
&& \ +  \frac{ v_F}{2\pi \lambda}\sum_\alpha \left\{ r_\alpha(t)
\exp\left[ i\alpha\hat{\phi}^{C}_{+}(0)+
i\hat{\phi}^{I}_{+}(0)\right] +
h. c.\right\},
\label{bosH}
\end{eqnarray}
where the complex function of time $z(t)$ is given by Eq.~(\ref{z}).
The bosonized current operator (\ref{current}) becomes
\begin{equation}
\hat{I}=\left.\frac{e v_F}{2\pi}\frac{\partial
\hat{\phi}^{I}_+}{\partial x}
\right| _{x=0}.
\label{bosI}
\end{equation}

As one can see from Eq.~(\ref{bosH}), the mode
$\phi_+^C(0)$ is pinned by the charging energy $E_C$ to the value $\pi
N$.
Since  $E_C$ is the large scale
in the problem, for the description of the low energy dynamics of
the system  we can integrate $\phi_+^C$ out\cite{Matveev95}
and obtain the Hamiltonian of the form
\begin{eqnarray}
\label{bosH2}
\hat{H}'(t)&=&\frac{v_F}{4\pi}
\int_{-\infty}^{+\infty}dx
\left( \frac{\partial \hat{\phi}^{I}_{+}(x)}{\partial x} \right)^2
\nonumber \\
& + & \sqrt{\frac{\gamma v_F E_C}{2\pi^3\lambda}} \left[ z(t)
\exp( i\hat{\phi}^{I}_{+}(0)) + h. c.\right],
\end{eqnarray}
where $\ln \gamma={\bf C}\approx 0.5772$ is the Euler constant.

The Hamiltonian (\ref{bosH2}) reduces to a non-interacting
form through the introduction of the fermion fields\cite{Matveev95}
\begin{equation}
\hat{\Psi}(x)=\frac{\hat{\zeta}}{\sqrt{2\pi\lambda}}
\exp\left(i \hat{\phi}^I_{+}(x) \right),
 \label{referm}
\end{equation}
where $\hat{\zeta}$ is a Majorana fermion.

The re-fermionized Hamiltonian (\ref{bosH2})
can be conveniently written in a matrix form
\begin{mathletters}
\begin{eqnarray}
\hat{H}_F(t)&=& \frac{1}{2}\int_{-\infty}^{+\infty} dx
\hat{\Upsilon}^\dagger(x)\hat{H}_0 \hat{\Upsilon}(x),\\
\hat{\Upsilon}^\dagger(x)&=& \left( \hat{\Psi}^\dagger(x),
\hat{\Psi}(x), \zeta \right),\\
\hat{H}_0 & =&
\pmatrix{
-iv_F \partial_x & 0 &  \kappa z^*(t)\delta (x)\cr
0   & iv_F \partial_x  &
-\kappa z(t)\delta (x) \cr
\kappa z(t)\delta (x)  & \kappa z^*(t)\delta (x)&  0\cr
},
  \label{fermH}
 \end{eqnarray}
where $\kappa= \sqrt{\gamma v_F E_c/\pi^2 } $.
The re-fermionized current operator (\ref{bosI}) acquires the form
\end{mathletters}
 \begin{equation}
\hat{I}_F= e v_F :\hat{\Psi}^\dagger(0) \hat{\Psi}(0):
   \label{fermI}
  \end{equation}

Next, we define
the matrix Green function $\hat{G}^<(t,t';x,x')$
which includes both normal and anomalous components
\begin{eqnarray}
\hat{G}(t,t';x,x')& =& - i \left\langle T_t
\hat{\Upsilon}(x,t)\otimes
\hat{\Upsilon}^\dagger (x',t')
\right\rangle
  \label{GFdef}
 \end{eqnarray}
where $\langle \ldots \rangle$ denotes averaging over the quantum state
of
the system. The retarded and advanced Green functions
$\hat{G}^{R/A}(t,t';y,y')$ are defined in a similar manner.

All of the observables can be expressed through  the Green function
at almost coinciding times, $\hat{{\cal G}}(t;x,x')$
\begin{equation}
\hat{{\cal G}}(t;x,x') =  \hat{G}(t,t+0;x,x') ,
\label{calGF}
 \end{equation}
where $\hat{G}(t,t';x,x')$ is defined in Eq.~(\ref{GFdef}).
For example, the instantaneous current through the system
is given by
 \begin{equation}
I(t)= - i ev_F \hat{{\cal G}}_{11}(t;x=0,x'=0).
  \label{Ivalue}
 \end{equation}

In the leading adiabatic approximation the Green functions of
the system coincide with those in equilibrium at
the instantaneous value of $z$, and the
current (\ref{Ivalue}) vanishes. Therefore, to find the
current flowing through the system in response to
an adiabatic change of the Hamiltonian we have to find the first
non-adiabatic correction to the Green function (\ref{calGF}).
It obeys the evolution equation
\begin{equation}
\frac{d \hat{\cal G}(t)}{dt}=- i
\left[\hat{H}_0 (t), \hat{\cal G}(t)\right],
\label{evolution}
\end{equation}
where $ \hat{H}_0 (t)$ is given by Eq.~(\ref{fermH}). From
this equation
it follows that  the first non-adiabatic correction to the
Green function (\ref{calGF}) is given by
\begin{equation}
\delta\hat{\cal G}(t)=-\int
\frac{d \omega}{2\pi}
\hat{G}^R_0\left[z(t),\omega\right] \frac{d\hat{\cal
G}_0\left[z(t)\right]}{dt}
\hat{G}^A_0\left[z(t),\omega\right],
 \label{GFcorr}
\end{equation}
where $\hat{\cal G}_0[z(t)]$ is the equilibrium Green function
(\ref{calGF}) for the instantaneous value of the parameter
$z(t)$, and $\hat{G}_0^{R/A}[z(t),\omega]$
are the frequency
representations for the retarded and advanced Green functions at
fixed $z(t)$. They can be found from the following equations
\begin{mathletters}
 \begin{eqnarray}
&&\left( \omega _\pm{\rm diag}\left[1,1,1/2 \right]
 - \hat{H}_0(x)\right)
\hat{G}_0^{R/A}(z(t),\omega;x,x')= \nonumber \\
&&\hspace*{1.5cm}{\rm diag}\left[ \delta(x-x'), \delta (x-x'),1\right],
  \label{GReqdef} \\
&& \hat{\cal G}_0\left[ z(t)\right]=  \int
\frac{d\omega}{2\pi}
n_F(\omega)
\left( \hat{G}_0^{A}(z,\omega)- \hat{G}_0^{R}(z,\omega)\right),
  \label{Geqdef}
 \end{eqnarray}
\label{GFseqdef}
where $n_F(\omega)=\left[1+\exp(\omega/T)\right]^{-1}$ is the Fermi
 distribution
function, and $\omega_\pm \equiv \omega \pm i0$.
\end{mathletters}

We solve Eq.~(\ref{GReqdef}) for the Green functions,
substitute the result in Eqs.~(\ref{Geqdef}) and
(\ref{GFcorr}), and thus find the current (\ref{Ivalue}).
Integrating
the result over the cycle period we find the charge transmitted
during the cycle. It can be represented through the
 the dimensionless conductance of the system $g$
(in units of $e^2/2\pi\hbar$) as

\begin{equation}
Q = \frac{1}{2\pi i}\oint \frac{dz}{z}
\left[1 - 2 g (|z|^2, T)\right],
\label{qfinal}
\end{equation}
where the dimensionless conductance is given by~\cite{Furusaki95}
\[
g=\frac{1}{2} - \frac{\gamma |z(t)|^2 E_c}{2 \pi^3 T}
\zeta\left(2,\frac{1}{2}+\frac{|z(t)|^2\gamma E_c}{\pi^3 T}\right),
\]
with  $\zeta\left(x,y\right)$ being the Riemann zeta-function.
At low temperatures, $T \ll |z|^2 E_c$, conductance vanishes as
$g \propto T^2/(|z|^2 E_c)^2$, which means that the transmitted charge
tends to its quantized value. At high temperatures
$T\gtrsim |z|^2 E_c$ the conductance approaches the classical value
$g=1/2$,
and the pumped charge (\ref{qfinal}) vanishes. In general,
for $g\neq 0$, the pumped
charge depends on the shape of the contour $z(t)$ and is not a
topological number.

We stress that Eq.~(\ref{qfinal}) for the transmitted charge
is given by the sum of two terms: i) the first one arises
from non-dissipative currents (This contribution is
quantized and represents a topological invariant of the cycle.),
and  ii) the second one, containing the conductance, is
due to dissipative currents generated by the cycling of the dot.

This fact is not accidental and becomes more transparent from
the following consideration which applies to a more general
class of systems. The time dependent Hamiltonian can be written as
$\hat{H}(t) = \hat{A}^\dagger(t) \hat{H}_0(t)\hat{A}(t)$, where
$\hat{H}_0$ is diagonal.
In the adiabatic limit the transmitted charge can be
most conveniently evaluated by going to the adiabatically
rotating basis $|\tilde{\psi}_i(t)\rangle=
\hat{A}(t)|\psi_i(t)\rangle$ (the
``rotating axis representation''\cite{Messiah}),
and calculating the current
in response to the arising perturbation
$i \partial_t \hat{A}(t)\hat{A}^\dagger(t)$
using the Kubo formula
\begin{equation}
\label{Kubo}
I=-\int_{-\infty}^0 dt \langle [ \partial_t
\hat{A}(t)\hat{A}^\dagger(t), \hat{I}(0)]\rangle.
\end{equation}

Now let us apply Eq.~(\ref{Kubo}) to the case of an open dot
which is connected to the leads by two groups of channels
denoted by index $\alpha_i = \pm 1$, and which is described
by the Hamiltonian similar to Eq.~(\ref{Heff}). For this
purpose we introduce the partial particle number operator
in each channel
\begin{equation}\label{chargeI}
\hat{n}_i= \int_{0}^{\infty}  dx
\psi^\dagger_i(\alpha_i x)\psi_i(\alpha_i x)
\end{equation}
We notice, that even though the average particle number
$\langle \hat{n}_i\rangle_{H}$
in each channel is infinite, its change during the pumping cycle
is a well defined quantity and is determined by the
the gate voltage, $N(t)$ and the reflection coefficients $r_j$.
The calculation is facilitated
by the explicit form of the unitary operator
\begin{equation}
\label{unitary}
\hat{A}(t)= \prod_j \exp
\left( \frac{i \pi \alpha_j \langle \hat{n}_j\rangle_{H}}{v_F}
 \int_{-\infty}^{+\infty}\! dx\  \hat{I}_i (x)\right),
\end{equation}
where  $\hat{I}_j (x)$
is the partial particle current operator in channel $j$ at point $x$,
\begin{equation}
\label{Ipart}
\hat{I}_j(x) = \frac{i}{2m}\left[\psi_j^\dagger\partial_x\psi_j-
\left(\partial_x \psi_j^\dagger\right)\psi_j\right].
\end{equation}
Recalling that the time evolution of operators at $x\neq 0$
corresponds to free propagation with velocity $v_F$ we readily
express the pumping current through the dimensionless
partial conductances $g_{ij}$ between the channels
\begin{eqnarray}\label{theorem}
I=-\frac{ e}{2} \sum_{i,j}\alpha_j
\frac{ d\langle \hat{n}_i\rangle_{H}}{dt}
(\delta_{ij}-g_{ij}), \nonumber \\
g_{ij} = 2\pi i  \int_{-\infty}^{0} dt \ t
\left[\hat{I}_i(t,0); \hat{I}_j(0,0)\right].
\end{eqnarray}

For the model of spinless electrons, Eq.~(\ref{theorem})
reproduces Eq.~(\ref{qfinal}),
since $g_{1,-1}$ is the two terminal conductance of the system,
$g_{1,1} = g_{1,-1}=g$ because of charge conservation, and
$\langle \hat{n}_1+\hat{n}_{-1}\rangle_H = - N(t)$, because of
the large charging energy. The explicit relation $\langle
\hat{n}_1-\hat{n}_{-1}\rangle_H = -\frac{1}{\pi}
{\rm Im} \ln z(t)$ with $z$ given by Eq.~(\ref{z}) requires
the model assumption (\ref{Heff}).

Equation (\ref{theorem}) holds even in the case
when the elastic cotunneling \cite{AverinNazarov}
can not be neglected. As shown in Ref.~\cite{Aleiner97} the
elastic returns in the strong tunneling case can be
described by an $N$-independent action. Therefore, the
explicit form of the unitary operator (\ref{unitary})
and Eq.~(\ref{theorem}) are still valid.
This observation gives us an estimate of
 $\Delta/(|z|^2 E_C)$ for
the sample-specific\cite{Aleiner96,Aleiner97}
correction to the quantized value of the charge
due to the finite level spacing in the dot, $\Delta$,
since  the conductance in the valley of
the CB remains finite even in the zero-temperature limit
$g \simeq \Delta/(|z|^2 E_C)$\cite{Aleiner97}.

Let us now apply Eq.~(\ref{theorem}) to strong inelastic cotunneling
 of electrons with spin.
Because of the spin symmetry
and charge conservation we have $\sum_s g_{\alpha,\alpha'}^{s,s'}=
g/2$, where $s$ is the spin index and $g$ is the two
terminal conductance, and we readily obtain
\begin{equation}
I =  \  -\frac{e}{2}\sum_{\alpha, s} \alpha
\frac{d \langle \hat{n}_{\alpha}^s\rangle_H
}{dt}
 \left[1 -  g(t) \right].
\label{qs}
\end{equation}
At high temperature  $g\to 1$, and pumping is suppressed,
whereas at low temperatures the conductance vanishes
as $T^2$~\cite{Furusaki95}. The explicit
relation between $\langle \hat{n}_{\alpha}^s\rangle_H$ and
the parameters of the Hamiltonian
is a difficult problem which was not solved. However, for the
anisotropic case, where the reflection coefficient in
one contact is much stronger than 
in the other, one can show that $\sum_{\alpha, s} \alpha
d \langle \hat{n}_{\alpha}^s\rangle_H
/dt \approx   \left( dN(t)/dt \right)
{\rm sgn}(\sum_\alpha \alpha |r_\alpha|) $.
Moreover, if in the transition region, $r_1 \sim r_{-1}$,
we do not go through a degeneracy point $\cos \pi N =0$
the pumped charge remains quantized at zero temperature, and the
problem is topologically equivalent to the spinless case.
All estimates of the finite level spacing
effect remain the same as for the spinless case.

To summarize, we considered adiabatic charge transport
through a quantum dot. We have
shown, that even in the case of small backscattering
in the channel, when quantum fluctuations of the dot charge
are large, the transmitted charge is still quantized.
We have calculated  corrections to the quantized value
of the charge due to finite temperature and
the level spacing in the dot. These  corrections
are expressed through the conductance $g$ and originate from
the dissipative currents generated by pumping.

We are grateful to B.L.~Altshuler,   A.~Ludwig,  C.M.~Marcus,
B.Z.~Spivak, and F.~Zhou for interesting discussions. I.A. is
an A.P.~Sloan research fellow. A.A. is
supported by NSF Grant No.~PHY94-07194.

\end{multicols}
\end{document}